\documentclass[traditabstract]{aachanged}
\usepackage[english]{babel}
\usepackage{amsmath}
\usepackage{amssymb}
\usepackage{euscript}
\usepackage{cyr}
\usepackage{epsfig}
\begin{document}

\title{Variation of the baryon-to-photon ratio \\
       due to decay of dark matter particles}
\titlerunning{Variation of the baryon-to-photon ratio}
\author{E. O. Zavarygin$^{1,2}$\thanks{E-mail: e.zavarygin@gmail.com} and A. V. Ivanchik$^{1,2}$\thanks{E-mail: iav@astro.ioffe.ru}}

\authorrunning{Zavarygin and Ivanchik}
\date{Received 05 December, 2014}

\institute{\it{$^{1}$Ioffe Institute, ul. Politekhnicheskaya 26, St. Petersburg, 194021 Russia} \\
\it{$^{2}$Peter the Great St.Petersburg Polytechnic University, ul. Politekhnicheskaya 29, St. Petersburg,
Russia}}

\abstract{The influence of dark matter particle decay on the baryon-to-photon ratio has been studied for different cosmological epochs.
We consider different parameter values of dark matter particles such as mass, lifetime, the relative fraction of dark matter particles. It is shown that the modern value of the dark matter density $\Omega_{\rm CDM}=0.26$ is enough to lead to variation of the baryon-to-photon ratio up to $\Delta \eta / \eta \sim 0.01 \div 1$ for decays of the particles with masses $10\,$GeV$\,\div\,1\,$TeV. However, such processes can also be accompanied by emergence of an excessive gamma ray flux. The observational data on the diffuse gamma ray background are used to making constraints on the dark matter decay models and on the maximum possible variation of the baryon-to-photon ratio $\Delta\eta/\eta\lesssim10^{-5}$. Detection of such variation of the baryon density in future cosmological experiments can serve as a powerful means of studying properties of dark matter particles.}


\maketitle

\section{INTRODUCTION}
\label{introduction}

In the last decade, cosmology has passed into the
category of precision sciences. Many cosmological
parameters are currently determined with a high precision
that occasionally reaches fractions of a percent
(Ade et al. 2014). One of such parameters is the
baryon-to-photon ratio $\eta \equiv n_{\rm b}/n_{\gamma}$, where $n_{\rm b}$ and $n_{\gamma}$
are the baryon and photon number densities in the
Universe, respectively. In the standard cosmological
model, the present value of $\eta$ is assumed to have
been formed upon completion of electron-positron
annihilation several seconds after the Big Bang and
has not changed up to now.

The value of $n_{\gamma}$ associated with the cosmic microwave
background (CMB) photons is defined by
the well-known relation
$$
n_{\gamma}=\frac{2\zeta(3)}{\pi^2}\left( \frac{kT}{\hbar c}\right)^3=410.73\left(\frac{T}{2.7255\,\text{K}}\right)^3\text{cm}^{-3}, 
$$
where $\zeta(x)$ is the Riemann zeta function, $k$ is the
Boltzmann constant, $\hbar$ is the Planck constant, $c$ is
the speed of light, and $T$ is the CMB temperature
at the corresponding epoch. The CMB temperature
is currently determined with a high accuracy and is
$T_0 = 2.7255(6)\,$K at the present epoch (Fixsen 2009);
for other epochs, it is expressed by the relation $T=T_0(1 + z)$, where $z$ is the cosmological redshift at
the corresponding epoch. Thus, given $n_{\gamma}$, a relation
between the parameter $\eta$ and $\Omega_{\rm b}$, the relative baryon
density in the Universe, can be obtained (Steigman 2006):
\begin{equation*}
\eta = 273.9\times10^{-10}\Omega_{\rm b}h^2,
\end{equation*}
where $h = 0.673(12)$ is the dimensionless Hubble
parameter at the present epoch (Ade et al. 2014).
According to present views, the baryon density,
which is the density of ordinary matter (atoms,
molecules, planets and stars, interstellar and intergalactic
gases), does not exceed 5\% of the entire
matter filling the Universe, while 95\% of the density
in the Universe is composed of unknown forms
of matter/energy that manifest themselves (for the
time being) gravitationally (see, e.g., Gorbunov and
Rubakov 2008).

At present, observations allow $\Omega_{\rm b}$ to be independently
estimated for four cosmological epochs:\\
(i) the epoch of Big Bang nucleosynthesis ($z_{\rm BBN}\sim10^9$; see, e.g., Steigman et al. 2007);\\
(ii) the epoch of primordial recombination ($z_{\rm PR}\simeq1100$; see, e.g., Ade et al. 2014);\\
(iii) the epoch associated with the Ly$\alpha$ forest ($z\sim2\div3$; i.e., $\sim$10~Gyr ago; see, e.g., Rauch 1998; Hui et al. 2002);\\
(iv) the present epoch ($z = 0$; see, e.g., Fukugita and Peebles 2004).

For the processes at the epochs of Big Bang nucleosynthesis
and primordial recombination, $\eta$ is one
of the key parameters determining their physics. For
these epochs, the methods of estimating $\eta$, (i) comparing
the observational data on the relative abundances
of the primordial light elements (D, $^4$He, $^7$Li)
with the predictions of the Big Bang nucleosynthesis
theory and (ii) analyzing the CMB anisotropy, give
the most accurate estimates of $\eta$ to date that coincide,
within the observational error limits: $\eta_{\rm BBN} = (6.0 \pm 
0.4) \times 10^{-10}$ (Steigman 2007) and $\eta_{\rm CMB} = (6.05 \pm
0.07) \times 10^{-10}$ (Ade et al. 2014). This argues for the
correctness of the adopted model of the Universe and
for the validity of the standard physics used in theoretical
calculations. However, it should be noted that
at present, as the accuracy of observations increases,
some discrepancy between the results of observations
and the abundances of the primordial elements predicted
in the Big Bang nucleosynthesis theory has
become evident. The ``lithium problem'' is well known
(see, e.g., Cyburt et al. 2008); not all is ideal with
helium and deuterium (for a detailed discussion of
these problems, see Ivanchik et al. 2015). These
inconsistencies can be related both to the systematic
and statistical errors of experiments and to the
manifestations of new physics (physics beyond the
standard model).

The determination of $\Omega_{\rm b}$ and the corresponding
$\eta$ at epochs (iii) and (iv) has a considerably lower
accuracy. The value of $\eta$ measured for the epoch
associated with the Ly$\alpha$ forest coincides, by an order of magnitude,
with $\eta_{\rm BBN}$ and $\eta_{\rm CMB}$, but, at the same
time, is also strongly model-dependent (e.\,g., Hui et al. 2002). The measured
$\Omega_{\rm b}$ and $\eta$ at the present epoch are at best half those
predicted by Big Bang nucleosynthesis calculations
and CMB anisotropy analysis. The so-called problem
of missing baryons (see, e.g., Nicastro et al. 2008) is
associated with this.

It is hoped that further observations and new experiments
will allow $\Omega_{\rm b}$ for different cosmological
epochs and the corresponding $\eta$ to be determined with
a higher accuracy. In turn, this can become a powerful
tool for investigating the physics beyond the standard
model, where the values of $\eta$ for different cosmological
epochs can be different. Constraints on the deviation
of $\eta$ allow various theoretical models admitting such
a change to be selected.

In this paper, we discuss the possibility of a change
in $\eta$ on cosmological time scales attributable to the
decays of dark matter particles. For example, supersymmetric
particles (see, e.g., Jungman et al. 1996;
Bertone et al. 2004; and references therein) can act
as such particles; some of them can decay into the
lightest stable supersymmetric particles and standard
model particles (baryons, leptons, photons, etc.; see,
e.g., Cirelli et al. 2011):
\begin{equation}
{\rm X} \rightarrow \chi + ... \begin{cases}
 {\gamma + \gamma +...} \\
 {\rm p + \bar{p} +...}, 
 \end{cases}
\end{equation} 
where X and $\chi$ are unstable and stable dark matter
particles, respectively. This can lead to a change in $\eta$.

The currently available observational data suggest
that the dark matter density in the Universe is approximately
a factor of 5 larger than the baryon density:
$\Omega_{\rm CDM}\simeq 5\Omega_{\rm b}$, i.e., the relation between the number
density of dark matter particles and the number
densities of baryons and photons in the Universe is
$n_{\rm CDM}\simeq 5(m_{\rm b}/m_{\rm CDM})n_{\rm b} = 5(m_{\rm b}/m_{\rm CDM})n_{\gamma}\eta$. Assuming
that the changes in the number densities of
various types of particles in the decay reactions of
dark matter particles are related as $\Delta n_{\rm CDM} \sim \Delta n_{\rm b}$
and $\Delta n_{\rm CDM} \sim \Delta n_{\gamma}$, it is easy to see that the parameter
$\eta$ is most sensitive precisely to the change in
baryon number density. In the decays of dark matter
particles with masses $m_{\rm CDM} \sim 10$\,GeV$-$1\,TeV,
the change in $\eta$ as a result of the change in baryon
number density could reach $\Delta\eta/\eta \sim 0.01 - 1$
\footnote{Here and below, out of all baryons, we restrict ourselves to
protons. This assumption is valid for obtaining estimates,
because the bulk of the baryon density in the Universe is
contained in the hydrogen nuclei, while heavier baryons (for
example, D, He, etc.) are generated with a considerably
lower probability.}. The
change in photon number density and the change
in $\eta$ attributable to it will be approximately billion
times smaller. Therefore, in our paper we focused our
attention on the possibility of a change in $\eta$ due to the
decays of dark matter particles with the formation of
a baryon component.

Despite the negligible contribution to the change
in $\eta$ from the photon component, a comparison of
the predicted gamma-ray background (dark matter
particle decay products) with the observed isotropic
gamma-ray background in the Universe can serve
as an additional source of constraints on the decay
models of dark matter particles. The photons
produced by such processes are high-energy ones.
The observational data on the isotropic gamma-ray
background constrain their possible number in the
Universe, which, in turn, narrows the range of admissible
parameters of dark matter particles, determines
the maximum possible number of baryons, the
decay products of dark matter particles, and the corresponding
change in the baryon-to-photon ratio in
such decays. Thus, the observational data on the
gamma-ray background, along with the cosmological
experiments described above, serve as a source 
of constraints on the decay models of dark matter
particles and on the possible change in $\eta$. Running
ahead, we will say that at present the constraints from
isotropic gamma-ray background observations are
more severe than those following from cosmological
experiments.

Depending on the lifetime of dark matter particles,
a statistically significant change in $\eta$ can occur
at different cosmological epochs. We consider
lifetimes $\tau$ in the following range: $t_{\rm BBN}\ll \tau \lesssim t_0$,
where $t_{\rm BBN}\simeq3\,$min is the age of the Universe at
the end of the epoch of Big Bang nucleosynthesis,
$t_0 \simeq 13.8\,$Gyr is the present age of the Universe (Ade
et al. 2014). The decays of dark matter particles with
short lifetimes ($\tau \lesssim t_{\rm BBN}$) can change significantly
the chemical composition of the Universe (see, e.g.,
Jedamzik 2004; Kawasaki et al. 2005). The available
observational data on the abundances of the primordial
light elements (D, $^4$He, $^7$Li) agree well with the
predictions of Big Bang nucleosynthesis calculations,
which, in turn, limits the possibility of such a change.
For long lifetimes exceeding the present age of the
Universe ($\tau > t_0$), the change in $\eta$ at the above four
cosmological epochs will be so small that this will
unlikely allow it to be detected without a significant
improvement in observational capabilities.

\section{THE BARYON-TO-PHOTON RATIO IN MODELS WITH PARTICLE DECAY}
\label{model}
\noindent
A large class of models with decaying dark matter
particles suggests the existence of the lightest stable
particle that we will designate as $\chi$. An unstable dark
matter particle, which we will designate as X, will
decay with time into a $\chi$-particle and standard model
particles. There can be reactions of the type  $\rm X \rightarrow \rm{\chi \, p \, \bar{p}}$
among such reactions, whose influence on $\eta$ is
investigated in this paper\footnote{Since we consider cosmological time scales, all of the neutrons
and antineutrons that are also produced in such decays
transform into protons and antiprotons.}. A quantitative parameter
characterizing the fraction of the decay channels of
X-particles whose products are hadrons (in our case,
these will be protons and antiprotons) in the total
number of decay channels is the hadronic branching
ratio $B_h$, which is $B_h = 1$ in our case.

The currently available observational data argue
for the absence (or a negligible amount) of relic antimatter
(baryon-asymmetric Universe). For this reason,
the parameter $\eta$ in the standard cosmological
model is defined as the ratio of the baryon number
density to the photon number density. Since in our
model the decays of X-particles will lead to the production
of protons and antiprotons, we will define the
parameter $\eta$ as the ratio of the sum of the baryon and
antibaryon number densities to the photon number
density in the Universe:
\begin{eqnarray}
\eta(z) = \frac{n_{\rm b}(z) + n_{\rm \bar{b}}(z)}{n_{\gamma}(z)}&&\\ \nonumber
= \frac{n_{\rm b}^{\rm BBN}(z) + \Delta n_{\rm p}(z) + \Delta n_{\rm \bar{p}}(z)}{n_{\gamma}^{\rm BBN}(z)} &=& \eta_{\rm BBN} + \Delta \eta (z),
\label{eta_definition}
\end{eqnarray}
where $n_{\rm b}^{\rm BBN}$ and $n^{\rm BBN}_{\gamma}$ are the baryon and photon
number densities corresponding to $\eta_{\rm BBN} = n^{\rm BBN}_{\rm b} /n^{\rm BBN}_{\gamma}$;
$\Delta n_{\rm p}(z)$ and $\Delta n_{\rm \bar{p}}(z)$ are the number densities of
X-particle decay products: protons and antiprotons,
respectively (in the model under consideration,
$\Delta n_{\rm p}(z) = \Delta n_{\rm \bar{p}}(z)$, i.e., the generated baryonic
charge is $\Delta B = 0$). It is this value of (2) that
would be measured when determining the speed of
sound of the baryon-photon plasma at the epoch
of CMB anisotropy formation in the case of proton
and antiproton generation in accordance with the
formula (see, e.g., Gorbunov and Rubakov 2010)
\begin{equation}
u_s^2=\frac{\delta p}{\delta \rho}=\frac{c^2}{3(1+3\rho_{\rm B\bar{\rm B}}/4\rho_{\gamma})},
\end{equation}
where $\rho_{\rm B\bar{\rm B}}=\rho_{\rm B}+\rho_{\bar{\rm B}}$ is the sum of the baryon and
antibaryon densities in the Universe. In the standard
cosmological model, this quantity coincides with the
baryon density of the Universe $\rho_{\rm B}$. Thus, the baryon-to-photon ratio determined when analyzing the CMB
anisotropy is also the ratio of the sum of the baryon
and antibaryon number densities to the photon number
density and has the following form in the presence
of X-particle decay products:
\begin{equation}
\eta_{\rm CMB} = \left. \frac{n_{\rm b}(z) + n_{\rm \bar{b}}(z)}{n_{\gamma}(z)}\right|_{z=z_{\rm PR}} = \eta_{\rm BBN} + \Delta \eta (z_{\rm PR}),
\end{equation}

Note that for very early decays the antiprotons
being produced have time to annihilate with protons,
and $\eta$ again returns to its initial value $\eta = \eta_{\rm BBN}$.
The decays of X-particles with long lifetimes will
occur in an already fairly expanded Universe; consequently,
the antiprotons being produced may not
have time to annihilate. Thus, the later $\eta$ can differ
from $\eta_{\rm BBN}$ and $\eta_{\rm CMB}$. However, during the formation
of a large-scale structure, when halos in which the
density of matter exceeds considerably the average
one is formed, an excess of antiprotons would lead to
enhanced gamma-ray radiation from them.

\section{INFLUENCE OF THE DECAY OF DARK MATTER PARTICLES ON THE CHANGE IN {\Large $\eta$}}
\label{Data}

\noindent
The evolution of the number densities of X-particles,
$\chi$-particles, protons, and antiprotons in the Universe is described by the system of kinetic
equations
\begin{align}
\label{NLSP_decay}
\frac{dn_{\rm X}}{dt} &+  3Hn_{\rm X} \;\,=\; -\Gamma n_{\rm X},\\
\label{LSP_ann}
\frac{dn_{\chi}}{dt} &+ 3Hn_{\chi} \;\;=\;  \Gamma n_{\rm X},\\
\label{proton_ann}
\frac{dn_{\rm p,\bar{p}}}{dt} &+ 3Hn_{\rm p,\bar{p}} =\; - \langle \sigma v\rangle^{\text{ann}}_{\rm p\bar{p}}n_{\rm p} n_{\rm \bar{p}} + B_h \Gamma n_{\rm X},
\end{align}
where Eq. (7) consists of two equations describing
the evolution of the proton and antiproton number
densities, $n_{\rm p}$ and $n_{\rm \bar{p}}$, respectively; $n_{\rm X}$ and $n_{\chi}$ are the
number densities of X- and $\chi$-particles, respectively;
$H = \dot{a}/a$ is the Hubble parameter; $a(t)$ is the scale
factor; $\Gamma = 1/\tau$ is the decay rate of X-particles;
$\langle \sigma v\rangle^{\text{ann}}_{\rm p\bar{p}}$ is the product of the relative velocity $v$ and
proton-antiproton annihilation cross section $\sigma_{\rm ann}$
averaged over the momentum with a distribution
function. In a wide energy range (10\,MeV $\lesssim T_{\rm \bar{p}} \lesssim$
10 GeV), this quantity may be considered a constant,
$\langle \sigma v\rangle^{\text{ann}}_{\rm p\bar{p}} = 10^{-15}\,$cm$^3$s$^{-1}$ (see, e.g., Stecker 1967;
Weniger et al. 2013). The parameters of the standard
cosmological model presented in Table~1 are used to
solve Eqs.~(5)--(7).

\begin{table}[h]
\caption{Cosmological parameters used in this paper}
\label{table1}
\vspace{2mm}
\begin{tabular}{c|c|c}
 \hline
{Parameter}&{Value}&{Reference$^1$}\\
\hline
$\Omega_{\text{R}}$   &	5.46$\times 10^{-5}$            & 1 \\
$\Omega_{\text{CDM}}$ & 0.265                           & 2 \\
$\Omega_{\text{b}}$   & 0.05                            & 2 \\
$\Omega_{\Lambda}$    & 0.685                           & 2 \\
$H_0$                 & 67.3 km\,s$^{-1}$\,Mpc$^{-1}$   & 2 \\
$t_0$                 & 13.8 Gyr                        & 2 \\
\hline
\multicolumn{2}{l}{}\\
\multicolumn{2}{l}{$^1$ 1 -- Fixsen (2009), 2 -- Ade et al. (2014)}\\
\end{tabular}
\end{table}

Apart from the decays of dark matter particles,
we investigated the processes of their annihilation.
We showed that the influence of the annihilation of
dark matter particles with an annihilation cross section
$\langle \sigma v\rangle^{\text{ann}}_{\chi\bar{\chi}} = 10^{-26}\,$cm$^3$s$^{-1}$ (see, e.g., Jungman
et al. 1996) in the case where the annihilation
products are protons and antiprotons, $\chi\bar{\chi} \rightarrow \rm p \bar{p}$, on
the change in $\eta$ on all the time scales of interest
could be neglected. This implies the absence of the
terms responsible for the annihilation of X- and $\chi$-particles in Eq.~(7). The change in $\eta$ attributable to
the annihilation of dark matter particles with masses
10\,GeV--1\,TeV alone is negligible even at the epoch
of Big Bang nucleosynthesis (at which the contribution
from the annihilation is maximal): $|\Delta\eta/\eta_{\rm BBN}|<10^{-13}\div10^{-11}$ (the upper limit corresponds to a lower $\chi$-particle mass).

To determine the initial conditions for Eqs.~(5)
and (6), we introduce a parameter $\alpha$ defining the
fraction (by the number of particles) of unstable dark
matter particles in the entire dark matter at the epoch
of Big Bang nucleosynthesis. For the range of lifetimes
$t_{\rm BBN} \ll \tau \lesssim t_0$ we consider, the entire dark
matter at the present epoch will be composed of stable
$\chi$-particles some of which ($\alpha$) were produced by the
decays of X-particles and some ($1-\alpha$) are the relic
ones, i.\,e., the $\chi$-particle mass determines the initial
conditions for the X-particles as well. The availability
of reliable data on the parameter $\eta$ at the epoch of
Big Bang nucleosynthesis allows $\eta_{\rm BBN}$ to be used to
determine the initial condition for Eq.~(7). Thus, when
solving the system of equations (5)--(7), we use the
following initial conditions:
\begin{equation*}
z^0=z_{\rm BBN}=10^9, \quad t^0=\frac{1}{2H(z_{\rm BBN})},
\end{equation*}
\begin{equation}
\quad n^0_{\rm p}=\eta_{\rm BBN} n_{\gamma}(z_{\rm BBN}), \quad
n^0_{\bar{\text{p}}}=0,
\end{equation}
\begin{equation*}
n^0_{\chi}=(1-\alpha)\frac{\Omega_{\rm CDM}\rho_{\rm c}}{m_{\chi}c^2}, \quad n^0_{\rm X}=\alpha\frac{\Omega_{\rm CDM}\rho_{\rm c}}{m_{\chi}c^2},
\end{equation*}

Let us write the system of equations (5)--(7) in a
comoving volume that changes with time as $\sim a^3$, i.\,e.,
$\sim(1 + z)^{-3}$:
\begin{align}
\label{NLSP_decay_Y}
\frac{dY_{\rm X}}{dt} & = -\Gamma Y_{\rm X},\\
\label{LSP_ann_Y}
\frac{dY_{\chi}}{dt} & =  \Gamma Y_{\rm X},\\
\label{proton_ann_Y}
\frac{dY_{\rm p,\bar{p}}}{dt} & = - \langle \sigma v\rangle^{\rm ann}_{\rm p\bar{p}}Y_{\rm p} Y_{\rm \bar{p}}(1+z)^3 + B_h \Gamma Y_{\rm X},
\end{align}
where $Y_{i}= n_{i}/(1+z)^3$ is the number density of the
ith type of particles in the comoving volume.

In such a form, Eqs.~(9) and (10) have obvious
analytical solutions that describe the evolution of the
number densities of X- and $\chi$-particles in the comoving
volume:
\begin{equation}
Y_{\rm X}(t) = Y_{\rm X}^0 e^{-t/\tau},
\label{NLSP_solution}
\end{equation}
\begin{equation}
Y_{\chi}(t) = Y_{\chi}^0 + Y_{\rm X}^0(1-e^{-t/\tau}),
\end{equation}
where $Y_{\rm X}^0=n^0_{\rm X}/(1+z^0)^3$ and $Y_{\chi}^0=n_{\chi}^0/(1+z^0)^3$ are
the initial number densities of X- and $\chi$-particles in
the comoving volume. Substituting solution (12),
$\Gamma = 1/\tau$, and $B_h = 1$ into Eq.~(11), we obtain the final
system of equations describing the evolution of the
proton and antiproton number densities in the model
under consideration:
\begin{equation}
\frac{dY_{\rm p,\bar{p}}}{dt}  = - \langle \sigma v\rangle^{\rm ann}_{\rm p\bar{p}}Y_{\rm p} Y_{\rm \bar{p}}(1+z)^3 + \frac{Y_{\rm X}^0}{\tau}e^{-t/\tau}.
\label{prot_antiprot_eq}
\end{equation}

The corresponding change in the baryon-to-photon
ratio,
\begin{equation}
\frac{\Delta\eta(z)}{\eta_{\rm BBN}}=\frac{\eta(z) - \eta_{\rm BBN}}{\eta_{\rm BBN}},
\end{equation}
determined from the solution of the system of equations
(14) for $m_{\chi} = 10\,$GeV, $\alpha = 0.5$, and various $\tau$ is
presented in Fig.~1a. Note that the parameters $\alpha$ and
$m_{\chi}$ enter into the system of equations (14) in the form
of a ratio. Therefore, the result presented in Fig.~1
also corresponds to the case of larger masses of dark
matter particles provided that $\alpha/m_{\chi}$ is conserved.

\begin{figure*}
\centering
\includegraphics[width=0.75\textwidth,clip]{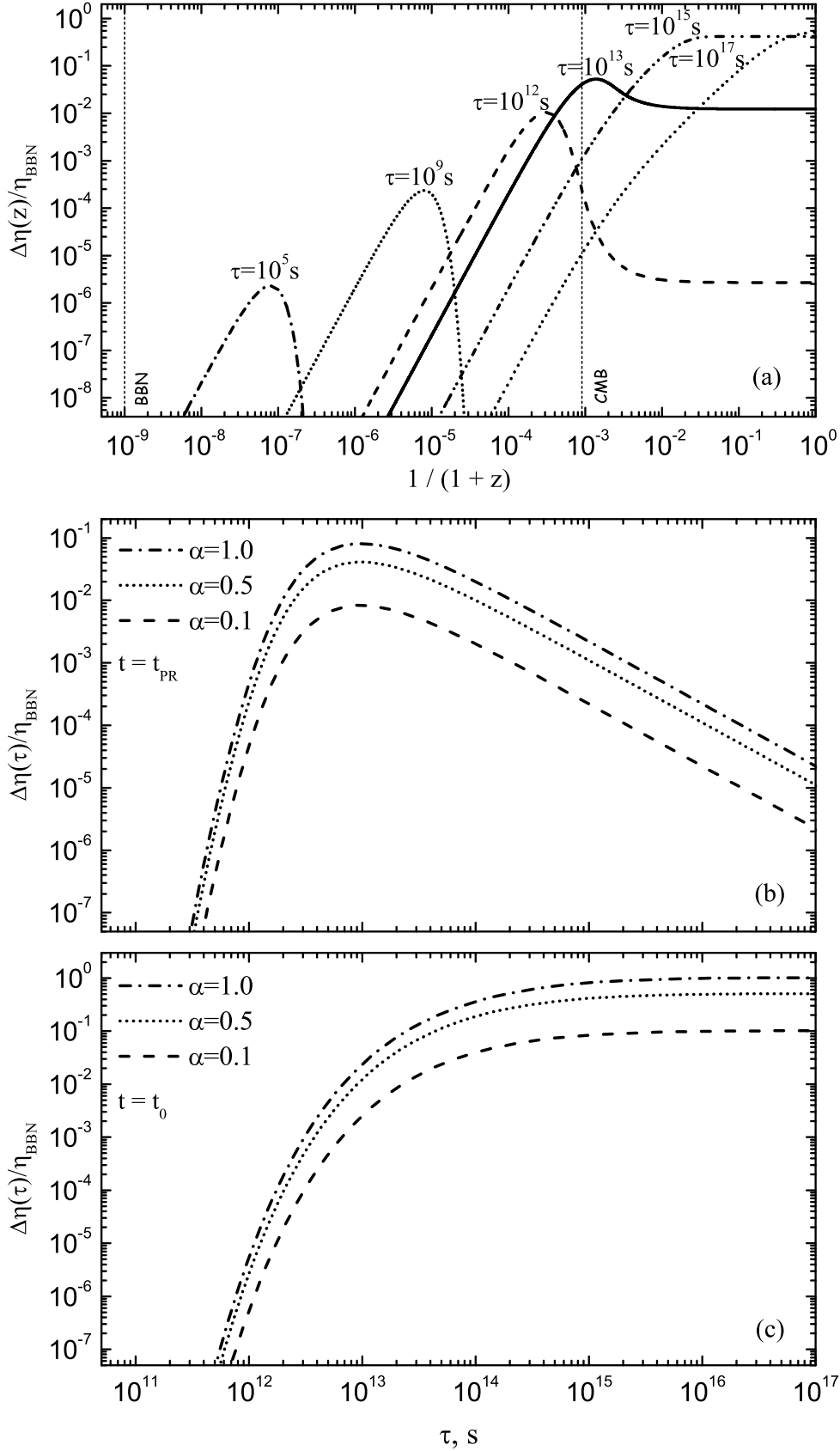}
 \caption{Fraction of the change in the baryon-to-photon ratio $\frac{\Delta\eta(z)}{\eta_{\rm BBN}}=\frac{\eta(z)-\eta_{\rm BBN}}{\eta_{\rm BBN}}$ attributable to the decays of X-particles
with lifetimes $10^5 {\,\rm s} \leq \tau \leq 10^{17}\,$s ($m_{\chi} = 10\,$GeV, $\alpha = 0.5$); the vertical lines mark the epochs of Big Bang nucleosynthesis ($z_{\rm BBN} \sim 10^9$) and primordial recombination ($z_{\rm PR} \simeq 1100$). The dependence $\Delta\eta(\tau)/\eta_{\rm BBN}$ of the change in the baryon-to-photon
ratio at the epoch of primordial recombination $t = t_{\rm PR}$ (b) and at the present epoch $t = t_0$ (c) on the lifetime of X-particles for various values of the parameter $\alpha$.}
 \label{eta_decay}
\end{figure*}

We see that the change in the baryon-to-photon
ratio in the model under consideration for lifetimes
$\tau \gtrsim 10^{12}\,$s can reach $\Delta\eta(z)/\eta_{\rm BBN} \sim 0.01-1$, which
is a potentially observable value. We also see that
the number densities of the protons and antiprotons
in the comoving volume produced by late decays ($\tau > 10^{13}\,$s) in an already fairly expanded Universe freeze
in such a way that $\eta$ can differ significantly from $\eta_{\rm BBN}$
and $\eta_{\rm CMB}$ by the present epoch. Note, however, that
in the decays $\rm X \rightarrow \chi p\bar{p}$ with the conservation of baryonic
charge (i.\,e., $\Delta n_{\rm p}(t) = \Delta n_{\rm \bar{p}}(t)$), $\Delta \eta/\eta_{\rm BBN} \sim 1$ at
the present epoch would imply almost equal numbers
of protons and antiprotons in the Universe, while
our Universe is significantly asymmetric in baryonic
charge. The existence of such a number of antiprotons
in the Universe would also give rise to an excess
of the gamma-ray background from the annihilation
of protons with antiprotons (see the next section).\\
\indent
Figure~1b presents the dependence $\Delta \eta(\tau)/\eta_{\rm BBN}$ of
the change in $\eta$ at the epoch of primordial recombination
(the epoch for which the parameter $\eta$ has been
measured most precisely to date) on the lifetime of
X-particles $\tau$ for various $\alpha$. We see that the fraction
of the change in $\eta$ at this epoch can reach $\Delta\eta/\eta_{\rm BBN} \sim 0.01-0.1$, which is also a potentially observable value.
Figure~1c present the dependence $\Delta\eta(\tau)/\eta_{\rm BBN}$ referring
to the present epoch ($t_0 \simeq 13.8\,$Gyr). We
see that the decay of X-particles in the model under
consideration leads to a significant change in the
present baryon density for $\tau > 10^{13}\,$s. However, the
accuracy of its determination at an epoch $z \sim 2-3$
and at the present epoch is still considerably lower
than that for the epochs of Big Bang nucleosynthesis
and primordial recombination.\\
\indent
The results obtained should not come into conflict
with other observational data:\\
(1) The decays with a predominance of hadronic
channels at early epochs $\tau \ll t_{\rm PR}$ can change significantly
the chemical composition of the Universe
(see, e.g., Jedamzik 2004; Kawasaki et al. 2005).
The available observational data on the abundances
of the primordial light elements (D, $^4$He, $^7$Li) agree
well with the predictions of Big Bang nucleosynthesis
calculations, which, in turn, limits the possibility of
such a change.\\
(2) The decays with $\tau \sim t_{\rm PR}$ can distort the CMB
spectrum and affect the angular CMB anisotropy
(see, e.g., Chen and Kamionkowski 2004; Chluba and
Sunyaev, 2012). Comparison with observational data
also allows the possible models to be constrained
severely.\\
(3) The hadronic decays with $\tau \gtrsim t_{\rm PR}$ can give rise
to an excess gamma-ray background from the annihilation
of produced antiprotons with background
protons and directly from the decays of X-particles
(see the next section).\\
\indent
In our case, we used data on the isotropic gamma-ray
background to obtain constraints on the decays of
particles with $t_{\rm PR} \lesssim \tau \lesssim t_0$, because a maximal effect
of change in the baryon-to-photon ratio is expected
for such lifetimes of X-particles (see Fig.~1). As
we will see, at present these constraints are more
significant than those that can be given by present-day
cosmological experiments.

\section{CONSTRAINT ON THE POSSIBLE CHANGE IN {\Large $\eta$} ASSOCIATED WITH THE OBSERVATION OF AN ISOTROPIC GAMMA-RAY BACKGROUND}
\label{gamma_flux}
\noindent
As was shown by Cirelli et al. (2011), apart from
protons and antiprotons, photons and leptons will
also be present among the end decay products of
dark matter particles, with their fraction exceeding
considerably the fraction of baryons even in the case
of $B_h = 1$ (i.\,e., when the decays completely run via
hadronic channels). The reason is that apart from
protons and antiprotons, mesons are produced in
the hadronization process, which contribute to the
photon and lepton components. In addition, the
appearance of an antiproton fraction in the Universe
will be accompanied by the formation of an additional
gamma-ray background from the annihilation
of proton-antiproton pairs. The main gamma-ray
background as a result of such a process will
arise from the decay of the $\pi^0$ meson produced by
the proton-antiproton annihilation (Stecker 1967;
Steigman 1976). Both these processes, which can
be represented schematically as
\begin{figure*}
\centering
\includegraphics[width=0.6\textwidth,clip]{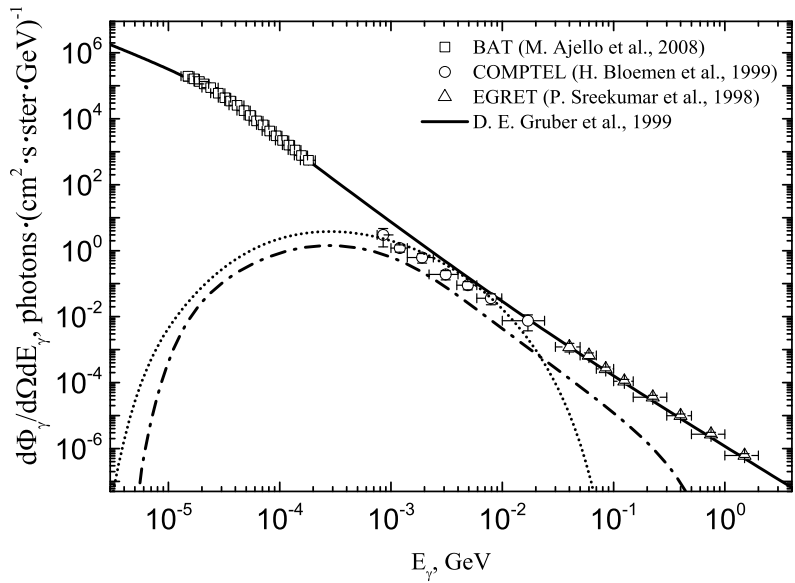}
 \caption{Isotropic gamma-ray background $d\Phi_{\gamma}/d\Omega dE_{\gamma}$ attributable directly to the decays of dark matter particles (dotted curve) and the annihilation of protons with antiprotons (dash-dotted curve) in the decay model of X-particles with a lifetime $\tau=10^{14}\,$s under consideration ($m_{\chi}=10\,$GeV, $m_{\rm X}-m_{\chi}=10\,$GeV, $\alpha=5\cdot10^{-6}$). The squares, circles, and triangles mark the experimental data taken from Ajello et al. (2008), Bloemen et al. (1999), and Sreekumar et al. (1998), respectively; the solid curve represents a fit to the experimental data from Gruber et al. (1999).}
 \label{flux for different species}
\end{figure*}
\begin{align}
{\rm X} \rightarrow \chi + ... \begin{cases}
 {\gamma + \gamma +...} \\
 {\rm p + \bar{p}} \rightarrow 
 \begin{cases}
   \pi^0     \;\, \rightarrow & \gamma + \gamma\\
   \pi^{\pm} \, \rightarrow & \mu^{\pm} + \nu_{\mu}(\tilde{\nu_{\mu}}),
 \end{cases}
 \end{cases} \\ \nonumber
 \mu^{\pm}\rightarrow e^{\pm} + \nu_{e}(\tilde{\nu_{e}}) + \nu_{\mu}(\tilde{\nu_{\mu}}),
\end{align}
will contribute to the isotropic gamma-ray background
in the Universe.

We calculate the corresponding gamma-ray background
by taking into account its extension to cosmological
distances. Note that photons of different
energies at different cosmological epochs interact
differently with the medium in which they propagate
(see, e.\,g., Zdziarski and Svensson 1989; Chen
and Kamionkowski 2004). More specifically, there
is a transparency window: the photons with energies
$E_{\gamma} < 10\,$GeV emitted at epochs $0 < z \lesssim 1000$
propagate almost without absorption and reach us
in the form of an isotropic gamma-ray background.
The formation of such a gamma-ray background is
expected from the decays of X-particles with lifetimes
$t_{\rm PR} \lesssim \tau \lesssim t_0$.

\begin{figure*}
\centering
\includegraphics[width=0.9\textwidth]{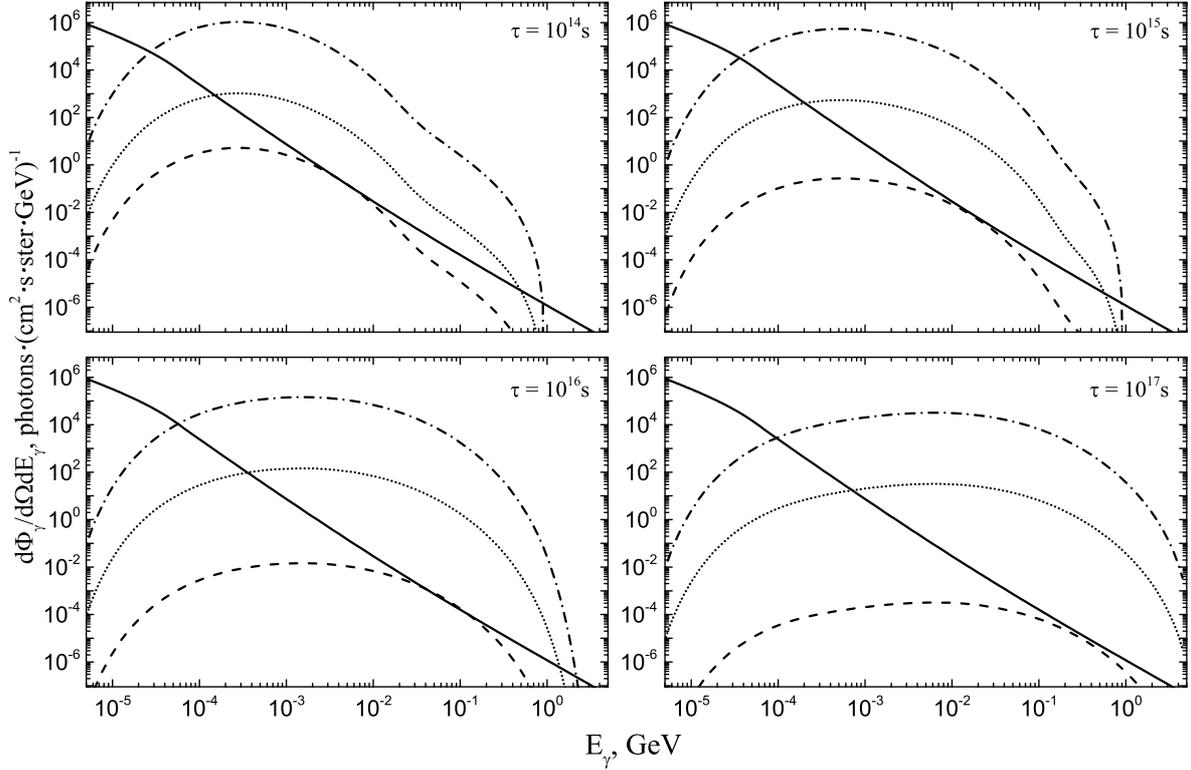}
\caption{Isotropic gamma-ray background $d\Phi_{\gamma}/d\Omega dE_{\gamma}$ in the decay model of X-particles with lifetimes $10^{14}\,$s~$\leq\tau\leq10^{17}\,$s
under consideration for $\alpha=1$ (dash-dotted curve), $\alpha=10^{-3}$ (dotted curve), and $\alpha=\alpha_{\rm max}$ (dashed curve) ($m_{\chi}=10\,$GeV, $m_{\rm X}-m_{\chi}=10\,$GeV). The solid curve represents a fit to the experimental data from Gruber et al. (1999).}
\label{diffuse flux alpha}
\end{figure*}

\begin{table*}
\caption{Maximum admissible fraction of X-particles $\alpha_{\text{max}}$ with various lifetimes $\tau$ for $\chi$-particle masses of 10, 100, and 1000~GeV and the corresponding maximum admissible change in the baryon-to-photon ratio $\Delta\eta/\eta_{\rm BBN}$ at epoch $z^*$.}
\begin{center}
\begin{tabular}{|c|c|c|c|c|c|}
\hline
           &  \multicolumn{3}{|c|}{$\alpha_{\rm{max}}$} &  &  \\
\cline{2-4}
 \raisebox{1.5ex}{$\tau,$ s} &  $m_{\chi}=10\,$GeV & $m_{\chi}=100\,$GeV & $m_{\chi}=1000\,$GeV & \raisebox{1.5ex}{$\frac{\Delta\eta(z^*)}{\eta_{\rm BBN}}$} & \raisebox{1.5ex}{$z^*$} \\
\hline
&&&&&\\
 $10^{14}$ & $5\times10^{-6}$ & $5\times10^{-5}$ & $5\times10^{-4}$ & $2.3\times10^{-6}$ & 120 \\
 $10^{15}$ & $5\times10^{-7}$ & $5\times10^{-6}$ & $5\times10^{-5}$ & $4.2\times10^{-7}$ & 18  \\
 $10^{16}$ & $10^{-7}$        & $10^{-6}$        & $10^{-5}$        & $10^{-7}$          & 2.2 \\
 $10^{17}$ & $10^{-8}$        & $10^{-7}$        & $10^{-6}$        & $10^{-8}$          & 0   \\
\hline
\end{tabular}
\label{table_alpha}
\end{center}
\end{table*}

The general formula describing the intensity of the
isotropic gamma-ray background $I_{\gamma}(E_{\gamma})$
(keV$\cdot$cm$^{-2}$s$^{-1}$sr$^{-1}$keV$^{-1}$) from various processes is
(see, e.\,g., Peacock 2010)
\begin{align}
&I_{\gamma}(E_{\gamma}) = E_{\gamma}\frac{d\Phi_{\gamma}}{d\Omega dE_{\gamma}}& \\ \nonumber
&= \frac{c}{4\pi}\int\limits_0^{1000} dz \frac{\epsilon_{\gamma}([1+z]E_{\gamma},z)}{H(z)(1+z)^4}e^{-\tau(E_{\gamma},z)},&
\label{extragalactic gamma flux general}
\end{align}
where $\Phi_{\gamma}$ is the gamma-ray photon flux per unit time
through a unit area, $\tau(E_{\gamma}, z)$ is the optical depth describing
the absorption of a photon emitted at epoch
$z$ with energy $E_{\gamma}(1 + z)$, $\epsilon_{\gamma}$ is the volume emissivity
that in our case is the sum of two terms,
\begin{equation}
\epsilon_{\gamma}(E_{\gamma},z) = \epsilon^{\rm X}_{\gamma}(E_{\gamma},z) + \epsilon^{\rm p\bar{p}}_{\gamma}(E_{\gamma},z),
\label{volume emissivity}
\end{equation}
describing the two contributions to the gamma-ray
background mentioned above. The first term $\epsilon^{\rm X}_{\gamma}$ is
related to the photons that are the X-particle decay
products; the second term $\epsilon^{\rm p\bar{p}}_{\gamma}$ is related to the photons
that are the proton-antiproton annihilation products.
These terms are described by the expressions
\begin{align}
\epsilon^{\rm X}_{\gamma}(E_{\gamma},z) = E_{\gamma} \Gamma n_{\rm X}(z) \frac{dN_{\gamma}}{dE_{\gamma}}\\\nonumber
= E_{\gamma} \Gamma Y_{\rm X}(z) (1+z)^3\frac{dN_{\gamma}}{dE_{\gamma}} ,
\label{volume emissivity X}
\end{align}
\begin{align}
\epsilon^{\rm p\bar{p}}_{\gamma}(E_{\gamma},z) = E_{\gamma} \langle \sigma v\rangle^{\rm ann}_{\rm p\bar{p}} n_{\rm p}(z) n_{\rm \bar{p}}(z)\frac{dN_{\gamma}}{dE_{\gamma}}\\\nonumber
 = E_{\gamma} \langle \sigma v\rangle^{\rm\ ann}_{\rm p\bar{p}} Y_{\rm p}(z) Y_{\rm \bar{p}}(z)(1+z)^6 \frac{dN_{\gamma}}{dE_{\gamma}},
\label{volume emissivity pp}
\end{align}
where $dN_{\gamma}/dE_{\gamma}$ is the spectrum of the photons
(phot$\cdot$keV$^{-1}$) emitted in one event of X-particle decay
(in Eq.~(19)) and proton-antiproton annihilation (in Eq.~(20)).

In our calculations, we use the spectrum $dN_{\gamma}/dE_{\gamma}$
of the photons that are the X-particle decay products
calculated in the PYTHIA package. The numerical
code for computing the spectra of the dark
matter particle decay and annihilation products was
taken from the site\footnote{http://www.marcocirelli.net/PPPC4DMID.html}; the details of using it can be
found in Cirelli et al. (2011). We use the data for
$m_{\rm X} - m_{\chi} \sim 10\,$GeV from the entire range of energy
release accessible in the numerical code in such
reactions, 10\,GeV$-$200\,TeV, to determine an upper
bound on the possible change in $\eta$. The optical
depths in (17) were also taken from this site.
The spectrum $dN_{\gamma}/dE_{\gamma}$ of the photons that are the
proton-antiproton annihilation products was taken
from Backenstoss et al. (1983).

\begin{figure*}
\centering
\includegraphics[width=0.9\textwidth,clip]{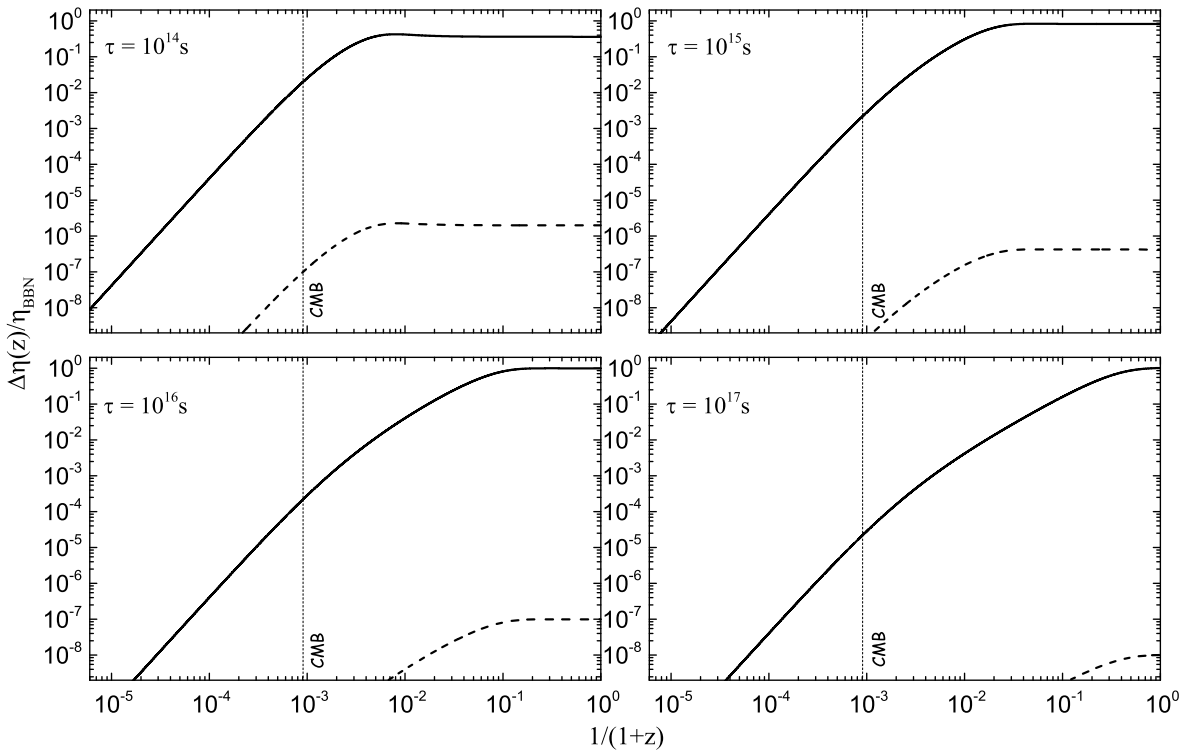}
 \caption{Fraction of the change in the baryon-to-photon ratio $\frac{\Delta\eta(z)}{\eta_{\rm BBN}}=\frac{\eta(z)-\eta_{\rm BBN}}{\eta_{\rm BBN}}$ attributable to the decays of X-particles
with lifetimes $10^{14}\,$s~$\leq\tau\leq10^{17}\,$s ($m_{\chi}=10\,$GeV, $m_{\rm X}-m_{\chi}=10\,$GeV). The solid and dashed curves correspond to the fractions of X-particles $\alpha=1$ and $\alpha=\alpha_{\rm max}$, respectively. The vertical line marks the epoch of primordial recombination ($z_{\rm PR}\simeq1100$).}
 \label{eta decay alpha max}
\end{figure*}

For comparison, Fig.~2 presents the gamma-ray
background $d\Phi_{\gamma}/d\Omega dE_{\gamma}$ (phot$\cdot$cm$^{-2}$s$^{-1}$sr$^{-1}$keV$^{-1}$)
attributable to the contribution from each of the two
terms in (18). The observational data on the isotropic
gamma-ray background (10\,keV$-$1\,GeV) taken from
Sreekumar et al. (1998), Bloemen et al. (1999),
Gruber et al. (1999), and Ajello et al. (2008) are also
presented in the figure. We see that the gamma-ray
background directly from the decays of X-particles
allow stringent constraints to be placed on the decay
processes.

Figure~3 shows the total gamma-ray background
$d\Phi_{\gamma}/d\Omega dE_{\gamma}$ with the inclusion of both terms in (18)
for X-particle lifetimes $t_{\rm PR} \lesssim \tau \lesssim t_0$ and various values
of the parameter $\alpha$. The gamma-ray background
admissible by the currently available observational
data corresponds to $\alpha_{\rm max}$, which characterizes the
maximum admissible fraction of unstable X-particles
with the corresponding lifetime. The values of $\alpha_{\rm max}$
for lifetimes $t_{\rm PR} \lesssim \tau \lesssim t_0$ are presented in Table~2.

Figure~4 presents the fraction of the change in
the baryon-to-photon ratio corresponding to $\alpha_{\rm max}$ for
various X-particle lifetimes (for comparison, Fig.~4
also presents this change for $\alpha = 1$). We see that this
change may reach $\Delta \eta(z)/\eta_{\rm BBN} \lesssim 10^{-5}$. The present-day
observational accuracy is $\Delta\eta/\eta \sim 10^{-2} - 10^{-1}$.
Note that the corresponding number of antiprotons in
the Universe at the present epoch related to $\Delta \eta$ via
the relation
\begin{equation*}
\frac{n_{\rm \bar{p}}}{n_{\rm p}}\simeq \left. \frac{1}{2}\frac{\Delta\eta}{\eta_{\rm BBN}}\right|_{z=0}
\end{equation*}
is consistent with the observational data on antiprotons
in cosmic rays (see, e.\,g., Adriani et al. 2010).

Since the parameters $\alpha$ and $m_{\chi}$ enter into the
system of equations (14) in the form of a ratio, the
result obtained can be easily generalized to the case
of larger masses of dark matter particles. For $\chi$-particles with masses $m_{\chi} = 10$, 100, and 1000~GeV,
the derived parameter $\alpha_{\rm max}$ and the corresponding
maximum change $\Delta\eta/\eta_{\rm BBN}$ in the baryon-to-photon
ratio are listed in Table~2. The table also gives the cosmological
redshift $z^{*}$ corresponding to the maximum
change in $\eta$.

\section{CONCLUSIONS}
\label{CONCLUSIONS}
\noindent
We investigated the influence of the baryonic decay
channels of dark matter particles $\rm X \rightarrow \chi p\bar{p}$ on
the change in the baryon-to-photon ratio at different
cosmological epochs.

We showed that the present dark matter density
$\Omega_{\rm CDM} \simeq 0.26$ is sufficient for the decay reactions of
dark matter particles with masses 10\,GeV$-$1\,TeV to
change the baryon-to-photon ration up to
$\Delta\eta(z)/\eta_{\rm BBN}\sim0.01-1$ (Fig~1). However, such a
change in $\eta$ would lead to an excess of the gamma-ray
background from the annihilation of proton-antiproton pairs, the decay products of dark matter
particles, and from the gamma-ray photons produced
directly in the decays of dark matter particles.

We used the observational data on the isotropic
gamma-ray background to constrain the decay models
of dark matter particles leading to a maximum
effect of change in $\eta$: we determined the maximum
admissible fraction of unstable dark matter particles
with lifetimes $t_{\rm PR} \lesssim \tau \lesssim t_0$ and the change in $\eta$ related
to them. The maximum possible change in the
baryon-to-photon ratio attributable to such decays is
$\Delta\eta(z)/\eta_{\rm BBN}\lesssim 10^{-5}$ (Fig.~4).

Despite the fact that at present the data on the
gamma-ray background constrain most severely the
decay models of dark matter particles with the emission
of baryons, the situation can change in future,
with increasing accuracy of existing cosmological
experiments and the appearance of new ones. The
detection of a change in the baryon-to-photon ratio
in such experiments at a level of $\lesssim10^{-5}$ will serve
as evidence for the existence of decaying dark matter
particles, while its detailed study will be a powerful
tool for studying their properties. In contrast, the
constancy of the baryon-to-photon ratio will serve as
a new source of constraints on the range of admissible
parameters of dark matter particles.

\section*{ACKNOWLEDGMENTS}
\label{ACKNOWLEDGMENTS}
\noindent
We thank the referees for their valuable remarks.
This work has been supported by the Russian Science Foundation (grant No~14-12-00955).

\end{document}